\documentclass[twocolumn,superscriptaddress,aps,pra]{revtex4-2}
\usepackage{amsfonts}
\usepackage{amsmath}
\usepackage{amssymb}
\usepackage{graphicx}
\usepackage{color}
\usepackage{ulem}
\usepackage{dcolumn}
\usepackage{bm}
\usepackage{url}
\usepackage[colorlinks, urlcolor = cyan, citecolor = blue, linkcolor = magenta]{hyperref}
\usepackage{ulem}
\usepackage{lipsum}
\usepackage{diagbox}
\usepackage{physics}
\def\be{\begin{equation}\begin{aligned}}
\def\ee{\end{aligned}\end{equation}}
\newcommand{\eq}[1]{\be #1 \ee}
\begin{document}

\title{Effective potential and scattering length of shielding polar molecules}
\author{Peng Xu}
\email{physicalxupeng@whu.edu.cn} 
\affiliation{School of Physics, Zhengzhou University, Zhengzhou 450001, China}
\affiliation{Institute of Quantum Materials and Physics, Henan Academy of Sciences, Zhengzhou 450046, China}
\author{Gang Chen}
\email{chengang971@163.com} 
\affiliation{Laboratory of Zhongyuan Light, School of Physics, Zhengzhou University, Zhengzhou 450001, China}

\begin{abstract}

    We investigate the effective potential and scattering length of ultracold polar molecules under different shielding techniques. First, we derive the effective potential for two polar molecules in the presence of an elliptical polarization field, combined elliptical and linear polarization fields, and combined elliptical polarization and static fields. The effective potential is then expressed as a sum of a zero-range contact interaction and a long-range dipole-dipole interaction under the Born approximation. We find that the first two shielding methods only partially suppress attractive interactions, while the second method allows for the construction of bound states with different polarization shapes. The last shielding method can achieve complete cancellation of residual attractive forces, which is particularly significant for maintaining quantum degeneracy in ultracold dipolar systems. Our results provide a comprehensive understanding of the effective potential and scattering length of shielding polar molecules, which is crucial for studying many-body physics in ultracold dipolar systems.

\end{abstract}

\maketitle

\section{Introduction}
\label{sec:intro}

Bose-Einstein condensates (BECs) have been a subject of intense research since their first experimental realization in alkali atoms in 1995~\cite{Anderson1995Observation, Davis1995Bose}. Since then, numerous intriguing phenomena have been observed in atomic BECs, such as quantized vortices~\cite{Abo2001Observation} and the superfluid-Mott-insulator transition~\cite{Greiner2002Quantum}. Atomic BECs also have wide applications in quantum simulations~\cite{Gross2017Quantum} and quantum metrology~\cite{Mao2023Quantum}. In recent years, polar molecules have attracted considerable attention due to their permanent electric dipole moments, which lead to long-range dipole-dipole interactions~\cite{Ni2008High, Ospelkaus2010Controlling, Takekoshi2014Ultracold, Molony2014Creation, Park2015Ultracold, Valtolina2020Dipolar, He2020Coherently}. Such interactions can give rise to novel quantum phases and applications that are challenging to achieve with short-range interactions. Examples include quantum droplets~\cite{Ferrier2016Observation}, supersolid phases~\cite{Casotti2024Observation}, Wigner crystals~\cite{Tsui2024Direct}, ultracold chemistry~\cite{De2011Controlling}, and quantum computation~\cite{Carroll2025Observation}.

However, polar molecules are highly susceptible to strong inelastic collisions~\cite{Ni2010Dipolar, De2011Controlling, Chotia2012Long, Stevenson2024Three}. Therefore, suppressing these collisional losses remains a critical challenge in stabilizing many-body dipolar molecules. Several methods have been proposed, such as optical lattice isolation~\cite{Chotia2012Long} and two-dimensional confinement using a strong electric field perpendicular to the plane~\cite{Ni2010Dipolar, De2011Controlling, Matsuda2020Resonant, Li2021Tuning}. However, the former approach cannot be used to study physics in continuous space, and the latter approach faces limitations in three-dimensional environments due to residual attractive forces along the head-to-tail molecular axis. Fortunately, shielding techniques provide an ideal approach to engineer repulsive barriers~\cite{Gorshkov2008Suppression, Karman2018Microwave, Lassab2018Controlling, Dutta2025Universality}. Utilizing these techniques, recent experimental breakthroughs have successfully realized stable ultracold polar molecular gases with high phase-space density~\cite{Lin2023Microwave, Bigagli2023Collisionally}, quantum degenerate Fermi gases~\cite{Schindewolf2022Evaporation}, and Bose-Einstein condensates~\cite{Bigagli2024Observation}.

Generally, multichannel scattering theory provides a fundamental framework for analyzing the collision dynamics of polar molecules~\cite{Taylor2012Scattering}. However, the complexity of multichannel scattering theory makes it difficult to study many-body physics. This methodological limitation motivates the development of an effective potential that captures the essential features of polar molecular interactions~\cite{Karman2022Resonant, Deng2023Effective, Deng2025Two, Karman2025Double}. Furthermore, the effective potential can be further simplified based on the Born approximation~\cite{Yi2000Trapped, Yi2001Trapped, Bohn2009Quasi}. The resulting formulation incorporates a zero-range contact interaction supplemented by a long-range dipole-dipole component. This combined potential enables systematic investigation of key quantum phenomena, including superfluidity, supersolidity, and quantum phase transitions at the mean-field level in dipolar systems.

In this paper, we systematically investigate the effective potential under three different shielding conditions: (I) an elliptical polarization field alone, (II) combined elliptical and linear polarization fields, and (III) combined elliptical polarization and static fields. We first derive the effective potential and calculate the scattering length for each shielding technique, comparing the results with those from the full Hamiltonian. Then, we simplify the effective potential as a zero-range contact interaction supplemented by a long-range dipole-dipole interaction under the Born approximation. Crucially, our analysis reveals that the first two shielding methods only partially suppress attractive interactions, potentially inducing three-body loss mechanisms. In contrast, the combination of elliptical polarization with static fields can achieve complete cancellation of residual attractive forces. This complete suppression mechanism is particularly significant for maintaining quantum degeneracy in ultracold dipolar systems.

\section{Hamiltonian of the system}
\label{sec:hamiltonian}

Here, we consider a gas of bosonic sodium-caesium (NaCs) molecules as a representative system~\cite{Bigagli2024Observation}, though the developed formalism applies broadly to quantum gases of both bosonic and fermionic species. At ultracold temperatures, the NaCs molecules remain in the $^{1}\Sigma(\nu = 0)$ rovibrational ground state with a permanent dipole moment of $d_0 = 4.75$ Debye. The resulting dipole-dipole interaction between two molecules is expressed as
\eq{
    V(\vb*{r}) &= \frac{d_0^2}{4 \pi \epsilon_0 r^3} \qty[ \vu*{d}_1 \cdot \vu*{d}_2 - 3 (\vu*{d}_1 \cdot \vu*{r}) (\vu*{d}_2 \cdot \vu*{r}) ] \\
    &= - 8 \sqrt{\frac{2}{15}} \pi^{3 / 2} \frac{d_0^2}{4 \pi \epsilon_0 r^3} \sum_{m = - 2}^{2} Y_{2, m}^*(\vu*{r}) \Sigma_{2, m},
}
where $\epsilon_0$ is the electric permittivity of vacuum, $\vu*{d}_j$ is the unit vector along the internuclear axis of molecule $j$, $r = \abs{\vb*{r}}$ is the distance between the two molecules, and $\vu*{r} = \vb*{r} / r$. The $\Sigma_{2, m}$ are rank-2 spherical tensors defined as $\Sigma_{2, 0} = \qty(Y_{1, 1}(d_1) Y_{1, -1}(d_2) + Y_{1, -1}(d_1) Y_{1, 1}(d_2) + 2 Y_{1, 0}(d_1) Y_{1, 0}(d_2)) \\ / \sqrt{6}$, $\Sigma_{2, \pm 1} = \qty(Y_{1, \pm 1}(d_1) Y_{1, 0}(d_2) + Y_{1, 0}(d_1) Y_{1, \pm 1}(d_2)) / \sqrt{2}$, and $\Sigma_{2, \pm 2} = Y_{1, \pm 1}(d_1) Y_{1, \pm 1}(d_2)$. Here, $Y_{l, m}$ are the spherical harmonics.

For the $j$-th molecule under shielding techniques, the internal Hamiltonian can be generally written as
\eq{
    H_{0, j} = b \vb{J}^2 + H_{ac, \sigma} + H_{ac, \pi} + H_{dc},
}
where $H_{ac, \sigma (\pi)} = - \vb*{d} \cdot \vb{E}_{\sigma (\pi)}$. Here, $\vb{E}_\sigma = E_\sigma e^{- i \omega_\sigma t} e^{i k_z z} \qty(\vu*{e}_+ \cos\alpha + \vu*{e}_{-} \sin\alpha) + \text{h.c.}$ represents an elliptically polarized field propagating along the $z$-axis, and $\vb{E}_\pi = E_\pi e^{- i \omega_\pi t} e^{i k_x x} \vu*{e}_z + \text{h.c.}$ represents a linearly polarized field propagating along the $x$-axis. $E_{\sigma (\pi)}$ is the amplitude, $\omega_{\sigma (\pi)}$ is the frequency of the microwave, and $\alpha$ is the ellipticity angle. Additionally, $\vu*{e}_{\pm} = \mp (\vu*{e}_x \pm i \vu*{e}_y) / \sqrt{2}$. Since the microwave wavelength is much larger than the molecular size and trap width, the alternating fields can be considered spatially uniform. The term $H_{dc} = - \vb*{d} \cdot \vb{E}_{dc}$ describes the interaction with a static field $\vb{E}_{dc} = E_{dc} \vu*{e}_z$ polarized along the $z$-axis. $\vb{J}$ is the total angular momentum operator of the molecule, and $b / \hbar = 2 \pi \times 1.74$ GHz is the rotational constant. Due to the anharmonicity of the rotational spectrum and for simplicity, we restrict to the two lowest rotational manifolds spanned by $\ket{J, m} = \ket{0, 0}, \ket{1, -1}, \ket{1, 0}, \ket{1, 1}$. The operator $Y_{1, m}(d)$ can then be expressed as $Y_{1, 0}(d) = (\ket{1, 0}\bra{0, 0} + \ket{0, 0}\bra{1, 0}) / \sqrt{4 \pi}$, $Y_{1, 1}(d) = (\ket{1, 1}\bra{0, 0} - \ket{0, 0}\bra{1, -1}) / \sqrt{4 \pi}$, and $Y_{1, -1}(d) = - Y_{1, 1}(d)^{\dagger}$.

Finally, the total Hamiltonian for the internal states of two molecules, $H = \sum_{j = 1, 2} H_{0, j} + V$, can be diagonalized in the 16-dimensional Hilbert space. However, the dipole-dipole interaction preserves parity symmetry, so the 10-dimensional even-parity subspace is decoupled from the remaining 6-dimensional odd-parity subspace. In the following, we focus on the even-parity subspace, which contains the relevant quantum states for studying effective interactions and scattering properties. By restricting to the even-parity subspace, we obtain an effective Hamiltonian that governs the system's low-energy dynamics. This approach significantly simplifies the theoretical analysis while preserving the essential physics of shielding dipolar interactions.

\section{Elliptical polarization field}
\label{sec:ellip}

In this section, we briefly review the effective potential of two polar molecules under an elliptical polarization field~\cite{Deng2023Effective}. Under the Born-Oppenheimer approximation, the Hamiltonian of two molecules can be written as
\eq{
    H = \sum_{j = 1, 2} H_{0, j} + V.
    \label{eq:total_ham_ellip}
}
Within the rotating-wave approximation, the Hamiltonian for a single molecule becomes
\eq{
    H_{0, j} = \delta \ket{0, 0} \bra{0, 0} + \frac{\Omega_\sigma}{2} \qty(\ket{\alpha_+}\bra{0, 0} + \text{h.c.}),
}
where $\delta = \omega_\sigma - 2 b$ is the detuning of the microwave, and $\Omega_\sigma = 2 d_0 E_\sigma / \sqrt{3}$ is the Rabi frequency. Here, $\ket{\alpha_+} = \cos\alpha \ket{1, 1} + \sin\alpha \ket{1, - 1}$ and $\ket{\alpha_-} = \cos\alpha \ket{1, - 1} - \sin\alpha \ket{1, 1}$. The eigenstates and their corresponding eigenvalues of the above Hamiltonian are $\ket{0} = \ket{1, 0}$ with $E_0 = 0$, $\ket{\alpha_-}$ with $E_{\alpha_-} = 0$, $\ket{-} = u \ket{0, 0} - v \ket{\alpha_+}$ with $E_{-} = (\delta - \Omega_{\text{eff}}) / 2$, and $\ket{+} = v \ket{0, 0} + u \ket{\alpha_+}$ with $E_{+} = (\delta + \Omega_{\text{eff}}) / 2$, where $\Omega_{\text{eff}} = \sqrt{\delta^2 + \Omega_\sigma^2}$, $u = \sqrt{(1 - \delta / \Omega_{\text{eff}}) / 2}$, and $v = \sqrt{(1 + \delta / \Omega_{\text{eff}}) / 2}$.

Besides, in the co-rotating frame, the rank-2 spherical tensor in the interaction term between two molecules can be approximately written as
\begin{widetext}
    \begin{align}
    &\Sigma_{2, 0} = \frac{1}{4 \pi \sqrt{6}} \big( 2 \ket{1, 0}\bra{0, 0} \otimes \ket{0, 0}\bra{1, 0} - \ket{1, 1}\bra{0, 0} \otimes \ket{0, 0}\bra{1, 1}
    - \ket{0, 0}\bra{1, -1} \otimes \ket{1, -1}\bra{0, 0} + \text{h.c.} \big), \nonumber \\
    &\Sigma_{2, 1} = \frac{1}{4 \pi \sqrt{2}} \big( \ket{1, 1}\bra{0, 0} \otimes \ket{0, 0}\bra{1, 0} - \ket{0, 0}\bra{1, -1} \otimes \ket{1, 0}\bra{0, 0} 
    + \ket{0, 0}\bra{1, 0} \otimes \ket{1, 1}\bra{0, 0} - \ket{1, 0}\bra{0, 0} \otimes \ket{0, 0}\bra{1, -1} \big), \nonumber \\
    &\Sigma_{2, -1} = - \Sigma_{2, 1}^{\dagger}, \nonumber \\
    &\Sigma_{2, 2} = - \frac{1}{4 \pi} \big( \ket{1, 1}\bra{0, 0} \otimes \ket{0, 0}\bra{1, -1} + \ket{0, 0}\bra{1, -1} \otimes \ket{1, 1}\bra{0, 0} \big), \nonumber \\
    &\Sigma_{2, -2} = \Sigma_{2, 2}^{\dagger},
    \label{eq:Sigma_2}
    \end{align}
\end{widetext}
where we have neglected the high-frequency terms. It turns out that the dipole-dipole interaction resides in the 7-dimensional Hilbert space spanned by $\ket{1} = \ket{+, +}$, $\ket{2} = \ket{+, 0}$, $\ket{3} = \ket{+, \alpha_-}$, $\ket{4} = \ket{+, -}$, $\ket{5} = \ket{-, 0}$, $\ket{6} = \ket{-, \alpha_-}$, $\ket{7} = \ket{-, -}$, with $\ket{\cdot, \cdot\cdot} = (\ket{\cdot, \cdot\cdot} + \ket{\cdot\cdot, \cdot}) / \sqrt{2}$. Correspondingly, the free energies of these states are $\delta + \Omega_{\text{eff}}$, $(\delta + \Omega_{\text{eff}}) / 2$, $(\delta + \Omega_{\text{eff}}) / 2$, $\delta$, $(\delta - \Omega_{\text{eff}}) / 2$, $(\delta - \Omega_{\text{eff}}) / 2$, $\delta - \Omega_{\text{eff}}$. The matrix form of the Hamiltonian in Eq.~\eqref{eq:total_ham_ellip} can be found in Appendix~\ref{sec:app1_ellip}.

The effective potential can be obtained by projecting the dipole-dipole interaction onto the system's highest-energy eigenstate manifold. Within the perturbative framework, the effective potential can be written as
\eq{
    V_{\text{eff}} &= \frac{C_3}{r^3} \qty[3 \cos^2 \theta - 1 + 3 \mathcal{F} \sin^2 \theta] \\
    &+ \frac{C_6}{r^6} \sin^2 \theta \qty[1 - \mathcal{F}^2 + \qty(1 - \mathcal{F})^2 \cos^2 \theta],
    \label{eq:eff_pot_ellip}
}
where $\mathcal{F} = \sin 2 \alpha \cos 2 \phi$, $C_3 = \frac{d_0^2}{48 \pi \epsilon_0} \qty(1 - \frac{\delta^2}{\Omega_\text{eff}^2})$, and $C_6 = \frac{d_0^4}{128 \pi^2 \epsilon_0^2 \Omega_\text{eff}} \qty(1 - \frac{\delta^2}{\Omega_\text{eff}^2})$. This result reproduces the essential features of earlier theoretical studies. The first term arises from first-order perturbation, representing the anti-dipolar interaction, which is repulsive along the $z$-axis and attractive in the $xy$-plane. The second term originates from second-order perturbation, including contributions only from the elements $\mel{1}{V}{2}$ and $\mel{1}{V}{3}$, as contributions from other elements can be safely neglected when $\delta_r = \delta / \Omega_\sigma \lesssim 1$. Notably, the repulsive core formed by the second-order term provides crucial stabilization against collapse dynamics.

\begin{figure}[t]
    \centering
    \includegraphics[width=8cm]{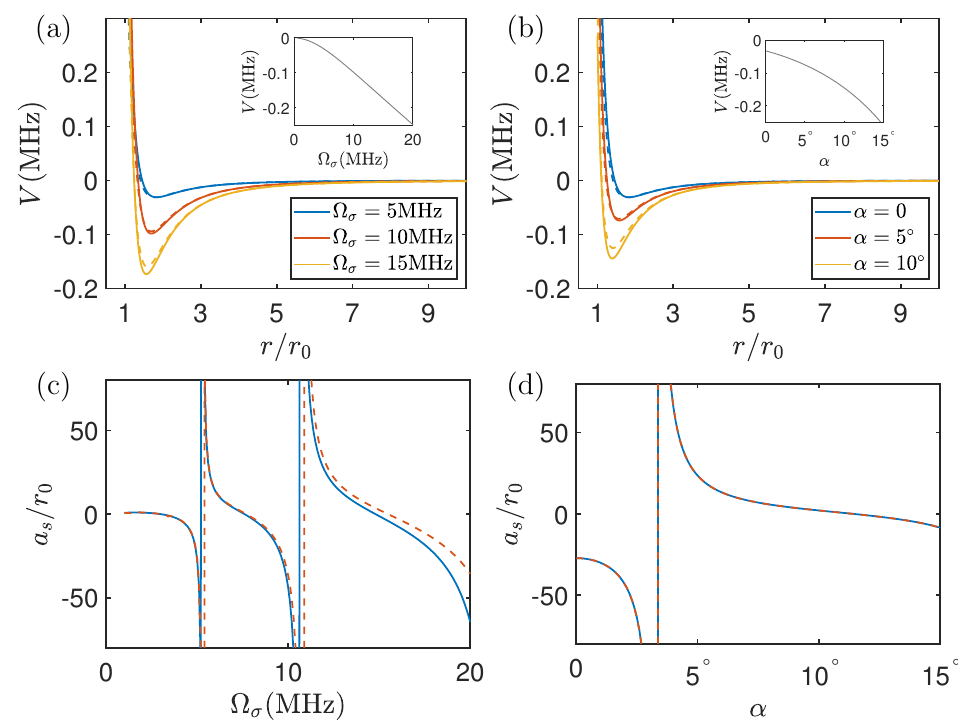}
    \caption{Adiabatic potentials along $\theta = \pi / 2$ for (a) $\delta = 10$ MHz, $\alpha = 0$, and $\Omega_\sigma = 5$, 10, 15 MHz in decending order; (b) $\delta = 10$ MHz, $\Omega_\sigma = 5$ MHz, and $\alpha = 0$, $5^\circ$, $10^\circ$ in decending order. The insets show the minimum of the effective potential for different control parameters. (c) Scattering length for $\delta = 10$ MHz, $\alpha = 0$, and different $\Omega_\sigma$; (d) $\delta = 10$ MHz, $\Omega_\sigma = 5$ MHz, and different $\alpha$. In (a)-(d), solid lines are obtained using the effective potential, and dashed lines are obtained using the full Hamiltonian. $r_0 = 10^3 a_B$, where $a_B$ is the Bohr radius.}
    \label{fig:eff_pot_ellip}
\end{figure}

In Fig.~\ref{fig:eff_pot_ellip} (a) and (b), we present the adiabatic potential curves for varying different parameters. A quantitative comparison confirms excellent agreement between the effective potential derived from perturbation theory and the exact diagonalization of the full Hamiltonian. The potential minimum decreases with increasing microwave intensity $\Omega_\sigma$ and ellipticity angle $\alpha$, indicating that more bound states may exist for larger $\Omega_\sigma$ and $\alpha$. Next, we calculate the scattering length $a_s$ for different parameters, as shown in Fig.~\ref{fig:eff_pot_ellip} (c) and (d). The detailed definitions and calculations of the scattering length for the effective potential and the full Hamiltonian can be found in Appendix~\ref{sec:app3_ellip} and Appendix~\ref{sec:app2}, respectively~\cite{Johnson1973Multichannel, Macedo2023Scattering}. We find that there are indeed more shape resonant states for larger $\Omega_\sigma$ and $\alpha$, and the scattering lengths obtained from the single-channel and multi-channel potentials are consistent, except for large $\Omega_\sigma$, because the single-channel effective potential curve shows more attractive characteristics.

After obtaining the scattering length, the effective potential can be further simplified as a zero-range contact interaction supplemented by a long-range dipole-dipole interaction under the Born approximation,
\eq{
    V_{\text{eff}}^{\text{Born}} &= \frac{2 \pi \hbar^2 a_s}{M} \delta(\vb*{r}) \\
    &+ \frac{C_3}{r^3} \qty[3 \cos^2 \theta - 1 + 3 \mathcal{F} \sin^2 \theta].
}
The validity of this pseudopotential framework is rigorously verified through partial wave analysis. Tables~\ref{tab:scat_len_ellip} and~\ref{tab:scat_len_ellip_1} compare the exact $t_{l m}^{l' m'}$ matrix elements with the pseudopotential predictions in Appendix~\ref{sec:app4_ellip}. We find that the ratio $t_{l m}^{l' m'} / t_{0 0}^{2 0}$ closely matches the values obtained using the pseudopotential in Appendix~\ref{sec:app4_ellip}, confirming the pseudopotential's quantitative accuracy within the Born approximation regime.

\begin{table}[h]
    \begin{tabular}{ccccc}
        \hline\hline
        $(l \; m), (l' m')$ & $(0 \; 0)$ & $(2 \; 0)$ & $(4 \; 0)$ & $(6 \; 0)$ \\
        \hline
        $(0 \; 0)$ & $*$ & $1$ & $0$ & $0$ \\
        $(2 \; 0)$ & $1$ & $- 0.6351$ & $0.1461$ & $0$ \\
        $(4 \; 0)$ & $0$ & $0.1461$ & $- 0.1713$ & $0.0596$ \\
        $(6 \; 0)$ & $0$ & $0$ & $0.0596$ & $- 0.0787$ \\
        \hline\hline
    \end{tabular}
    \caption{The ratio $t_{l m}^{l' m'} / t_{0 0}^{2 0}$ for different $l, m, l', m'$. $\delta = 10$MHz, $\Omega_\sigma = 5$MHz, $\alpha = 0$.}
    \label{tab:scat_len_ellip}   
\end{table}
\begin{table}[h]
    \begin{tabular}{ccccc}
        \hline\hline
        $(l \; m), (l' m')$ & $(0 \quad \; \; 0)$ & $(2 \quad \; \; 0)$ & $(2 \; - 2)$ & $(2 \quad \; \; 2)$ \\
        \hline
        $(0 \quad \; \; 0)$ & $*$ & $1$ & $- 0.2136$ & $- 0.2136$ \\
        $(2 \quad \; \; 0)$ & $1$ & $- 0.6349$ & $- 0.1347$ & $- 0.1347$ \\
        $(2 \; - 2)$ & $- 0.2136$ & $- 0.1347$ & $0.6344$ & $0$ \\
        $(2 \quad \; \; 2)$ & $- 0.2136$ & $- 0.1347$ & $0$ & $0.6344$ \\
        \hline\hline
    \end{tabular}
    \caption{The ratio $t_{l m}^{l' m'} / t_{0 0}^{2 0}$ for different $l, m, l', m'$. $\delta = 10$MHz, $\Omega_\sigma = 5$MHz, $\alpha = 5^\circ$.}
    \label{tab:scat_len_ellip_1}
\end{table}

\section{Combined elliptical and linear polarization fields}
\label{sec:ellip_pi}

To conveniently tune the effective potential, we can also consider the case of combining elliptical and linear polarization fields. Within the rotating-wave approximation, the Hamiltonian for a single molecule under the combined fields is given by
\eq{
    H_{0, j} &= \delta \ket{0, 0}\bra{0, 0} + \frac{\Omega_\sigma}{2} \qty(\ket{\alpha_+}\bra{0, 0} + \text{h.c.}) \\
    &+ \frac{\Omega_\pi}{2} \qty(\ket{1, 0}\bra{0, 0} + \text{h.c.}),
}
where $\Omega_{\sigma (\pi)} = 2 d_0 E_{\sigma (\pi)} / \sqrt{3}$, $\ket{\alpha_+} = \cos\alpha \ket{1, 1} + \sin\alpha \ket{1, - 1}$, and $\ket{\alpha_-} = \cos\alpha \ket{1, - 1} - \sin\alpha \ket{1, 1}$. For simplicity, we set the detuning $\delta = \omega_\sigma - 2 b = \omega_\pi - 2 b$. The eigenstates and their corresponding eigenvalues of the above Hamiltonian are $\ket{0} = p \ket{1, 0} - q \ket{\alpha_+}$ with $E_0 = 0$, $\ket{\alpha_-}$ with $E_{\alpha_-} = 0$, $\ket{-} = u \ket{0, 0} - v p \ket{\alpha_+} - v q \ket{1, 0}$ with $E_{-} = (\delta - \Omega_{\text{eff}}) / 2$, and $\ket{+} = v \ket{0, 0} + u p \ket{\alpha_+} + u q \ket{1, 0}$ with $E_{+} = (\delta + \Omega_{\text{eff}}) / 2$, where $\Omega_{\text{eff}} = \sqrt{\delta^2 + \Omega_t^2}$, $\Omega_t = \sqrt{\Omega_\sigma^2 + \Omega_\pi^2}$, $u = \sqrt{(1 - \delta / \Omega_{\text{eff}}) / 2}$, $v = \sqrt{(1 + \delta / \Omega_{\text{eff}}) / 2}$, $p = \Omega_\sigma / \Omega_t$, and $q = \Omega_\pi / \Omega_t$. Additionally, since the detuning of the $\sigma$ and $\pi$ microwaves is the same, the rank-2 spherical tensor in the co-rotating frame is the same as in Eq.~\eqref{eq:Sigma_2}. The dipole-dipole interaction also resides in the seven-dimensional Hilbert space. The matrix form of the Hamiltonian can be found in Appendix~\ref{sec:app1_ellip_pi}.

\begin{figure}[t]
    \centering
    \includegraphics[width=8cm]{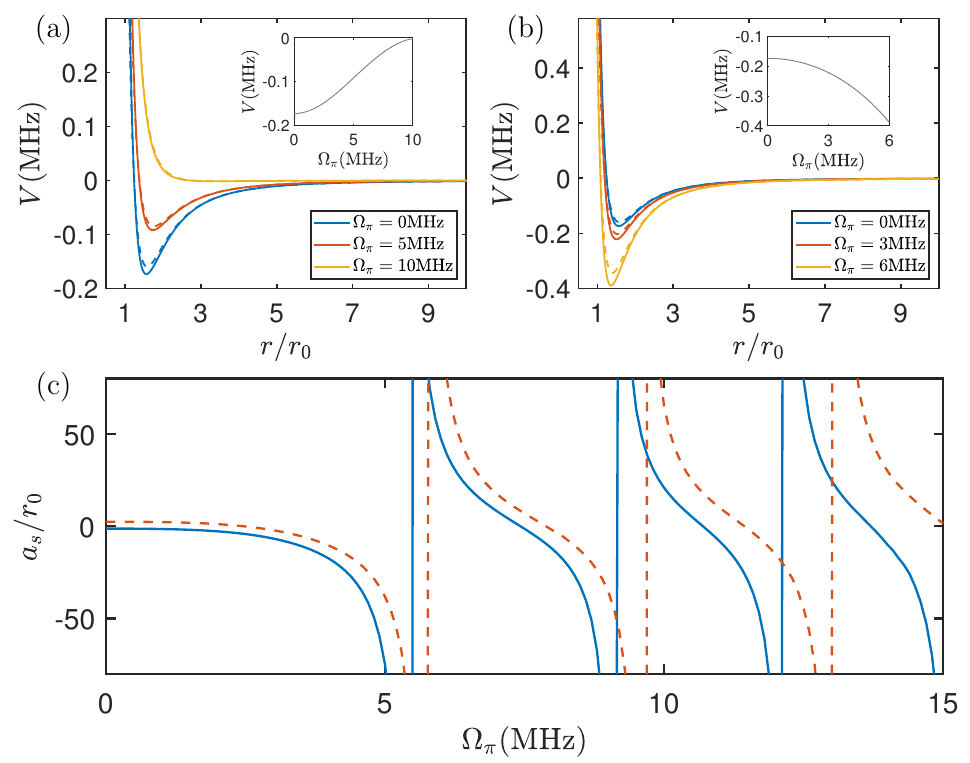}
    \caption{Adiabatic potentials along $\theta = \pi / 2$ for $\Omega_\pi = 0$, 5, 10 MHz in ascending order in (a); $\theta = \arccos(\sqrt{2} s / \sqrt{1 + 2 s^2})$ for $\Omega_\pi = 0$, 3, 6 MHz in descending order in (b). The insets show the minimum of the effective potential for different control parameters. (c) The scattering length for different $\Omega_\pi$. In (a)-(c), the solid lines are obtained using the effective potential, and the dashed lines are obtained using the full Hamiltonian. $r_0 = 10^3 a_B$, where $a_B$ is the Bohr radius. The other parameters are fixed as $\delta = 10$ MHz, $\Omega_\sigma = 15$ MHz, and $\alpha = 0$.}
    \label{fig:eff_pot_ellip_pi}
\end{figure}

Then the effective potential can be obtained based on perturbation theory,
\eq{
    V_{\text{eff}} &= \frac{C_3}{r^3} \qty[\qty(1 - 2 s^2) \qty(3 \cos^2 \theta - 1)] \\ 
    &+  \frac{C_3}{r^3} \qty[3 \mathcal{F} \sin^2 \theta + 3 \sqrt{2} s \mathcal{G} \sin 2 \theta] \\
    &+ \frac{C_6}{r^6} \qty[c_1 + p^2 c_2],
    \label{eq:eff_pot_ellip_pi}
}
where $\mathcal{F} = \sin 2 \alpha \cos 2 \phi$, $\mathcal{G} = (\cos \alpha - \sin \alpha) \cos \phi$, $s = \frac{\Omega_\pi}{\Omega_\sigma}$, $C_3 = \frac{d_0^2}{48 \pi \epsilon_0} \qty(1 - \frac{\delta^2}{\Omega_{\text{eff}}^2}) \frac{\Omega_\sigma^2}{\Omega_t^2}$, $C_6 = \frac{d_0^4}{128 \pi^2 \epsilon_0^2 \Omega_{\text{eff}}} \qty(1 - \frac{\delta^2}{\Omega_{\text{eff}}^2}) \frac{\Omega_\sigma^2}{\Omega_t^2}$. The expressions for $c_1$ and $c_2$ can be found in Appendix~\ref{sec:app1_ellip_pi}. When the linear polarization field is absent, i.e., $\Omega_\pi = 0$, the above effective potential reduces to the result in Eq.~\eqref{eq:eff_pot_ellip}. According to the first line of the above effective potential, we find that the anti-dipolar interaction is still attractive in the $xy$-plane when $\Omega_\pi$ is small, while the attractive strength decreases and even changes sign, becoming repulsive in the $xy$-plane as $\Omega_\pi$ increases, as shown in Fig.~\ref{fig:eff_pot_ellip_pi}(a). However, this does not imply that the effective potential becomes entirely repulsive or that there are no shallow bound states when $1 - 2 s^2 = 0$, as the most attractive direction is not in the $xy$-plane and changes for different $\Omega_\pi$ due to the second line of the effective potential. For example, when $\alpha = 0, \phi = 0$, the most attractive direction is along $\cos\theta \propto \sqrt{2} s / \sqrt{1 + 2 s^2}$, as shown in Fig.~\ref{fig:eff_pot_ellip_pi}(b), which allows us to tune the shape of the tetramer states in real space~\cite{Chen2023Field, Chen2024Ultracold}. The third line of the effective potential is always repulsive, providing a shielding core. The results for the scattering length $a_s$ are shown in Fig.~\ref{fig:eff_pot_ellip_pi}(c), and the detailed calculations can be found in Appendix~\ref{sec:app3_ellip_pi}. We find that the results are almost the same for the single-channel and multi-channel potentials for small $\Omega_\pi$. However, they deviate from each other for large $\Omega_\pi$, and the single-channel shows more resonant states, which is because the effective potential curve also shows more attractive characteristics as $\Omega_\pi$ increases. The effective potential and scattering length have similar results for $\alpha \neq 0$.

Finally, based on the Born approximation, the pseudopotential can be written as
\eq{
    V_{\text{eff}}^{\text{Born}} &= \frac{2 \pi \hbar^2 a_s}{M} \delta(\vb*{r}) \\
    &+ \frac{C_3}{r^3} \qty[\qty(1 - 2 s^2) \qty(3 \cos^2 \theta - 1)] \\
    &+ \frac{C_3}{r^3} \qty[3 \mathcal{F} \sin^2 \theta + 3 \sqrt{2} s \mathcal{G} \sin 2 \theta].
}
Tables~\ref{tab:scat_len_ellip_pi} and~\ref{tab:scat_len_ellip_pi_1} compare the exact $t_{l m}^{l' m'}$ matrix elements with the pseudopotential predictions from Appendix~\ref{sec:app4_ellip_pi}. We find that the ratio $t_{l m}^{l' m'} / t_{0 0}^{2 0}$ closely matches the values obtained using the pseudopotential in Appendix~\ref{sec:app4_ellip_pi}.

\begin{table}[h]
    \begin{tabular}{ccccc}
        \hline\hline
        $(l \; m), (l' m')$ & $(0 \; 0)$ & $(2 \; 0)$ & $(2 \; 1)$ & $(2 \; 2)$ \\
        \hline
        $(0 \; 0)$ & $*$ & $1$ & $- 0.3768$ & $0$ \\
        $(2 \; 0)$ & $1$ & $- 0.6391$ & $0.1210$ & $0$ \\
        $(2 \; 1)$ & $- 0.3768$ & $0.1210$ & $- 0.3182$ & $0.2938$ \\
        $(2 \; 2)$ & $0$ & $0$ & $0.2938$ & $0.6378$ \\
        \hline\hline
    \end{tabular} 
    \caption{The ratio $t_{l m}^{l' m'} / t_{0 0}^{2 0}$ for different $l, m, l', m'$. $\alpha = 0$, $\Omega_\sigma = 15$ MHz, $\Omega_\pi = 3$ MHz.}
    \label{tab:scat_len_ellip_pi}  
    \end{table}
    
\begin{table}[h]
    \begin{tabular}{ccccc}
        \hline\hline
        $(l \; m), (l' m')$ & $(0 \; 0)$ & $(2 \; 0)$ & $(2 \; 1)$ & $(2 \; 2)$ \\
        \hline
        $(0 \; 0)$ & $*$ & $1$ & $- 2.1620$ & $0$ \\
        $(2 \; 0)$ & $1$ & $- 0.6395$ & $0.6886$ & $0$ \\
        $(2 \; 1)$ & $- 2.1620$ & $0.6886$ & $- 0.3232$ & $1.6808$ \\
        $(2 \; 2)$ & $0$ & $0$ & $1.6808$ & $0.6429$ \\
        \hline\hline
    \end{tabular}
    \caption{The ratio $t_{l m}^{l' m'} / t_{0 0}^{2 0}$ for different $l, m, l', m'$. $\alpha = 0$, $\Omega_\sigma = 15$ MHz, $\Omega_\pi = 8$ MHz.}
    \label{tab:scat_len_ellip_pi_1}
\end{table}

\section{Combined elliptical polarization and static fields}
\label{sec:ellip_static}

\begin{figure}[t]
    \centering
    \includegraphics[width=8cm]{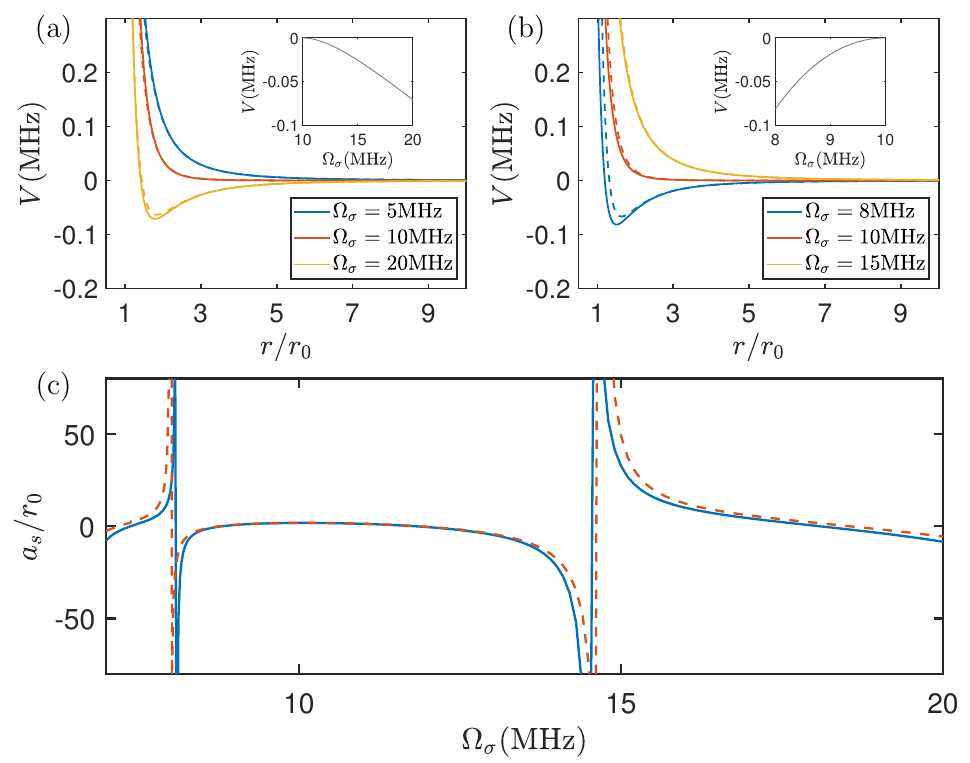}
    \caption{Adiabatic potentials along $\theta = \pi / 2$ for $\Omega_\sigma = 0$, 5, 10 MHz in descending order in (a); $\theta = 0$ for $\Omega_\sigma = 8$, 10, 15 MHz in ascending order in (b). The insets show the minimum of the effective potential for different control parameters. The anti-dipolar interaction is totally cancelled for $\delta = 10$ MHz and $\Omega_\sigma = 10$ MHz. (c) Scattering length for different $\Omega_\sigma$. In (a)-(c), the solid lines are obtained using the effective potential, and the dashed lines are obtained using the full Hamiltonian. $r_0 = 10^3 a_B$, where $a_B$ is the Bohr radius. The other parameters are fixed as $\delta = 10$ MHz, $\Omega_{dc} = 0.87 b$, and $\alpha = 0$.}
    \label{fig:eff_pot_ellip_static}
\end{figure}

Based on recent experimental results and the theoretical calculations in previous sections, at least one bound state exists for NaCs molecules under shielding techniques when $\Omega_\pi \sim \Omega_\sigma \sim \delta \sim 10$ MHz. This arises from the finite anti-dipolar interaction, which induces pronounced field-linked resonances (i.e., tetramer states) in the scattering process, significantly affecting the condensation lifetime~\cite{Bigagli2024Observation}. To improve the shielding technique and prolong the condensation lifetime, one may consider mismatched frequencies $\omega_\sigma \neq \omega_\pi$ when combining elliptical and linear polarization fields~\cite{Bigagli2024Observation, Deng2025Two, Karman2025Double}. Alternatively, one can also consider combining elliptical polarization and static fields~\cite{Gorshkov2008Suppression, Matsuda2020Resonant, Li2021Tuning}. The former method can be referenced in previous works~\cite{Deng2025Two, Karman2025Double}. Here, we systematically investigate the effective potential and scattering length for the hybrid elliptical polarization–static field configuration. Under the rotating-wave approximation, the single-molecule Hamiltonian under combined fields becomes
\eq{
    \hat{H}_{0, j} = \delta \ket{\beta_-}\bra{\beta_-} - \frac{\Omega_\sigma}{2} q \qty(\ket{\alpha_+}\bra{\beta_-} + \text{h.c.}),
}
where $\delta = \omega_\sigma - (b + \Omega_t / 2)$, $\Omega_t = \sqrt{(2 b)^2 + \Omega_{dc}^2}$, $\Omega_{\sigma (dc)} = 2 d_0 E_{\sigma (dc)} / \sqrt{3}$, $p = \sqrt{(1 - 2 b / \Omega_t) / 2}$, $q = \sqrt{(1 + 2 b / \Omega_t) / 2}$, $\ket{\alpha_+} = \cos\alpha \ket{1, 1} + \sin\alpha \ket{1, - 1}$, $\ket{\alpha_-} = \cos\alpha \ket{1, - 1} - \sin\alpha \ket{1, 1}$, $\ket{\beta_+} = q \ket{1, 0} + p \ket{0, 0}$, and $\ket{\beta_-} = p \ket{1, 0} - q \ket{0, 0}$. To ensure the validity of the rotating-wave approximation, we require $b \sim \Omega_{dc} \gg \Omega_\sigma$ and $\omega_\sigma \sim b + \Omega_t / 2$. The eigenstates and their corresponding eigenvalues of the above Hamiltonian are $\ket{\alpha_-}$ with $E_{\alpha_-} = 0$, $\ket{-} = u \ket{\beta_-} - v \ket{\alpha_+}$ with $E_- = (\delta - \Omega_{\text{eff}}) / 2$, and $\ket{+} = v \ket{\beta_-} + u \ket{\alpha_+}$ with $E_+ = (\delta + \Omega_{\text{eff}}) / 2$, where $\Omega_{\text{eff}} = \sqrt{\delta^2 + (\Omega_\sigma q)^2}$, $u = \sqrt{(1 - \delta / \Omega_{\text{eff}}) / 2}$, and $v = \sqrt{(1 + \delta / \Omega_{\text{eff}}) / 2}$.

In the co-rotating frame, the rank-2 spherical tensor in the interaction term between two molecules can be written as
\begin{widetext}
\eq{
    &\Sigma_{2, 0} = - \frac{q^2}{4 \pi \sqrt{6}} \big(\ket{1, 1}\bra{\beta_-} \otimes \ket{\beta_-}\bra{1, 1} + \ket{\beta_-}\bra{1, -1} \otimes \ket{1, -1}\bra{\beta_-}
    - 4 p^2 \ket{\beta_-}\bra{\beta_-} \otimes \ket{\beta_-}\bra{\beta_-} + \text{h.c.}\big), \\
    &\Sigma_{2, 1} = \Sigma_{2, -1} = 0, \\
    &\Sigma_{2, 2} = - \frac{q^2}{4 \pi} \big(\ket{1, 1}\bra{\beta_-} \otimes \ket{\beta_-}\bra{1, -1} + \ket{\beta_-}\bra{1, -1} \otimes \ket{1, 1}\bra{\beta_-}\big), \\
    &\Sigma_{2, -2} = \Sigma_{2, 2}^\dagger,
}
\end{widetext}
where we have also neglected the high-frequency terms. It turns out that the dipole-dipole interaction resides in the 5-dimensional symmetrized Hilbert space spanned by $\ket{1} = \ket{+, +}$, $\ket{2} = \ket{+, \alpha_-}$, $\ket{3} = \ket{+, -}$, $\ket{4} = \ket{-, \alpha_-}$, and $\ket{5} = \ket{-, -}$. The corresponding free energies of these states are $\delta + \Omega_{\text{eff}}, (\delta + \Omega_{\text{eff}}) / 2, \delta, (\delta - \Omega_{\text{eff}}) / 2, \delta - \Omega_{\text{eff}}$. The matrix form of the Hamiltonian can be found in Appendix~\ref{sec:app1_ellip_static}.

Then the effective potential can be calculated based on perturbation theory,
\eq{
    V_{\text{eff}} &= \frac{C_3}{r^3} \qty[\qty(1 - 4 s^2 p^2) \qty(3 \cos^2 \theta - 1) + 3 \mathcal{F} \sin^2 \theta] \\
    &+ \frac{C_6}{r^6} \qty[c_1 + \frac{u^4}{9} c_2],
    \label{eq:eff_pot_ellip_static}
}
where $\mathcal{F} = \sin 2 \alpha \cos 2 \phi$, $s = v / u$, $C_3 = \frac{d_0^2}{96 \pi \epsilon_0} \qty(1 - \frac{\delta^2}{\Omega_{\text{eff}}^2}) \qty(1 + \frac{2 b}{\Omega_t})$, and $C_6 = \frac{d_0^4}{512 \pi^2 \epsilon_0^2 \Omega_{\text{eff}}} \qty(1 - \frac{\delta^2}{\Omega_{\text{eff}}^2}) \qty(1 + \frac{2 b}{\Omega_t})^2$. The coefficients $c_1$ and $c_2$ are given in Appendix~\ref{sec:app1_ellip_static}. When the elliptical polarization field is absent, i.e., $\Omega_\sigma = 0$, the above effective potential is proportional to $\qty(1 - 3 \cos^2 \theta) / r^3$, which corresponds to the dipole-dipole interaction under full static polarization. Importantly, according to the above effective potential, we find that the anti-dipolar interaction can be completely canceled under the conditions $1 - 4 s^2 p^2 = 0$ and $\alpha = n \pi / 2$, $n \in \mathbb{Z}$, while the other terms provide a shielding core. This means the effective potential becomes fully repulsive, and there are no shallow bound states. This will significantly enhance the lifetime of the condensate. Interestingly, in a wide range $\Omega_\sigma \in (9, 14)$, there are also no bound states, such that the existing finite anti-dipolar interaction may not influence the condensation lifetime. The molecular system with finite but large anti-dipolar interaction, which is comparable to the $s$-wave scattering length, is significantly different from the atomic condensation platform. The numerical results for the effective potential are shown in Fig.~\ref{fig:eff_pot_ellip_static}.

We finally write the pseudopotential based on the Born approximation as
\eq{
    V_{\text{eff}}^{\text{Born}} &= \frac{2 \pi \hbar^2 a_s}{M} \delta(\vb*{r}) \\
    &+ \frac{C_3}{r^3} \qty[\qty(1 - 4 s^2 p^2) \qty(3 \cos^2 \theta - 1) + 3 \mathcal{F} \sin^2 \theta]. 
}
When the condition $1 - 4 s^2 p^2 = 0$ is satisfied, the effective potential becomes fully repulsive and the dipole-dipole interaction is canceled. Thus, the couplings between different partial waves vanish. Here, we consider the parameters $\alpha = 0$, $\delta = 10$ MHz, and $\Omega_\sigma = 12$ MHz such that there is a finite dipole-dipole interaction but no bound states. Table~\ref{tab:scat_len_ellip_static} compares the exact $t_{l m}^{l' m'}$ matrix elements with the pseudopotential predictions from Appendix~\ref{sec:app4_ellip_static}. The pseudopotential predictions are consistent with the exact results.

\begin{table}[h]
    \begin{tabular}{ccccc}
        \hline\hline
        $(l \; m), (l' m')$ & $(0 \; 0)$ & $(2 \; 0)$ & $(4 \; 0)$ & $(6 \; 0)$ \\
        \hline
        $(0 \; 0)$ & $*$ & $1$ & $0$ & $0$ \\
        $(2 \; 0)$ & $1$ & $- 0.6301$ & $0.1454$ & $0$ \\
        $(4 \; 0)$ & $0$ & $0.1454$ & $- 0.1705$ & $0.0594$ \\
        $(6 \; 0)$ & $0$ & $0$ & $0.0594$ & $- 0.0783$ \\
        \hline\hline
    \end{tabular}
    \caption{The ratio $t_{l m}^{l' m'} / t_{0 0}^{2 0}$ for different $l, m, l', m'$. $\alpha = 0$, $\delta = 10$MHz, $\Omega_\sigma = 12$MHz.}
    \label{tab:scat_len_ellip_static}   
\end{table}

\section{Conclusion}
\label{sec:conclusion and outlook}

In summary, we have systematically investigated the effective potential and scattering length of two molecules under different shielding techniques. The effective potential curves obtained by second-order perturbation theory are consistent with the results from the full Hamiltonian. Thus, the effective potential can be used to study the many-body physics of ultracold molecules, such as the ground state and excitation spectrum. After obtaining the scattering length, the effective potential can be further simplified as a zero-range contact interaction supplemented by a long-range dipole-dipole interaction under the Born approximation. The pseudopotential framework is rigorously verified through partial wave analysis. However, we note that the pseudopotential framework may not be valid in the negative scattering length regime, since a large shielding core exists. In the future, we will study the many-body physics, such as the ground state, quench dynamics, and Floquet dynamics of ultracold molecules, based on the effective potential and pseudopotential framework.

{\bf Acknowledgements} This project is supported by the National Natural Science Foundation of China (Grant No. 12375023 and No. 12204428), the Natural Science Foundation of Henan Province (Grant No. 242300421159), the National Key R \& D Program of China (Grants No. 2022YFA1404500 and No. 2021YFA1400900), the Cross-Disciplinary Innovative Research Group Project of Henan Province (Grant No. 232300421004), and the National Natural Science Foundation of China (Grant No. 12125406).

\appendix

\begin{widetext}
\section{Matrix form of the Hamiltonian for two molecules}
\label{sec:app1}

\subsection{Elliptical polarization field}
\label{sec:app1_ellip}

The Hamiltonian for two molecules under an elliptical polarization field is
\eq{
    H = \mathcal{E} - 8 \sqrt{\frac{2}{15}} \pi^{3 / 2} \frac{d_0^2}{4 \pi \epsilon_0 r^3} \sum_{m = - 2}^{2} Y_{2, m}^*(\vu*{r}) \Sigma_{2, m},
}
where 
\eq{\mathcal{E} = 
    \begin{pmatrix}
        \delta + \Omega_{\text{eff}} & 0 & 0 & 0 & 0 & 0 & 0 \\
        0 & \frac{1}{2} (\delta + \Omega_{\text{eff}}) & 0 & 0 & 0 & 0 & 0 \\
        0 & 0 & \frac{1}{2} (\delta + \Omega_{\text{eff}}) & 0 & 0 & 0 & 0 \\
        0 & 0 & 0 & \delta & 0 & 0 & 0 \\
        0 & 0 & 0 & 0 & \frac{1}{2} (\delta - \Omega_{\text{eff}}) & 0 & 0 \\
        0 & 0 & 0 & 0 & 0 & \frac{1}{2} (\delta - \Omega_{\text{eff}}) & 0 \\
        0 & 0 & 0 & 0 & 0 & 0 & \delta - \Omega_{\text{eff}}
    \end{pmatrix},
}
\eq{\Sigma_{2, 0} = \frac{1}{4 \pi \sqrt{6}}
    \begin{pmatrix}
        -2 u^2 v^2 & 0 & 0 & \sqrt{2} u v w & 0 & 0 & 2 u^2 v^2 \\
        0 & 2 v^2 & 0 & 0 & 2 u v & 0 & 0 \\
        0 & 0 & -v^2 & 0 & 0 & -u v & 0 \\
        \sqrt{2} u v w & 0 & 0 & - w^2 & 0 & 0 & - \sqrt{2} u v w \\
        0 & 2 u v & 0 & 0 & 2 u^2 & 0 & 0 \\
        0 & 0 & -u v & 0 & 0 & -u^2 & 0 \\
        2 u^2 v^2 & 0 & 0 & - \sqrt{2} u v w & 0 & 0 & -2 u^2 v^2 
    \end{pmatrix},
}
\eq{
    \Sigma_{2, 1} = \frac{1}{4 \pi \sqrt{2}}
    \begin{pmatrix}
        0 & \sqrt{2} u v^2 \cos\alpha & 0 & 0 & \sqrt{2} u^2 v \cos\alpha & 0 & 0 \\
        - \sqrt{2} u v^2 \sin\alpha & 0 & - v^2 \cos\alpha & v w \sin\alpha & 0 & - u v \cos\alpha & \sqrt{2} u v^2 \sin\alpha \\
        0 & - v^2 \sin\alpha & 0 & 0 & - u v \sin\alpha & 0 & 0 \\
        0 & - v w \cos\alpha & 0 & 0 & - u w \cos\alpha & 0 & 0 \\
        - \sqrt{2} u^2 v \sin\alpha & 0 & - u v \cos\alpha & u w \sin\alpha & 0 & - u^2 \cos\alpha & \sqrt{2} u^2 v \sin\alpha \\
        0 & - u v \sin\alpha & 0 & 0 & - u^2 \sin\alpha & 0 & 0 \\
        0 & - \sqrt{2} u v^2 \cos\alpha & 0 & 0 & - \sqrt{2} u^2 v \cos\alpha & 0 & 0
    \end{pmatrix},
}
\eq{
    \Sigma_{2, 2} = - \frac{1}{4 \pi}
    \begin{pmatrix}
        -u^2 v^2 \sin 2 \alpha &  & - \sqrt{2} u v^2 \cos^2 \alpha & \frac{\sqrt{2}}{2} u v w \sin 2 \alpha & 0 & -\sqrt{2} u^2 v \cos^2 \alpha & u^2 v^2 \sin 2 \alpha \\
        0 & 0 & 0 & 0 & 0 & 0 & 0 \\
        \sqrt{2} u v^2 \sin^2 \alpha & 0 & \frac{1}{2} v^2 \sin 2 \alpha & - v w \sin^2 \alpha & 0 & \frac{1}{2} u v \sin 2 \alpha & - \sqrt{2} u v^2 \sin^2 \alpha \\
        \frac{\sqrt{2}}{2} u v w \sin 2 \alpha & 0 & v w \cos^2 \alpha & - \frac{1}{2} w^2 \sin 2 \alpha & 0 & u w \cos^2 \alpha & - \frac{\sqrt{2}}{2} u v w \sin 2 \alpha \\
        0 & 0 & 0 & 0 & 0 & 0 & 0 \\
        \sqrt{2} u^2 v \sin^2 \alpha & 0 & \frac{1}{2} u v \sin 2 \alpha & - u w \sin^2 \alpha & 0 & \frac{1}{2} u^2 \sin 2 \alpha & - \sqrt{2} u^2 v \sin^2 \alpha \\
        u^2 v^2 \sin 2 \alpha & 0 & \sqrt{2} u v^2 \cos^2 \alpha & - \frac{\sqrt{2}}{2} u v w \sin 2 \alpha & 0 & \sqrt{2} u^2 v \cos^2 \alpha & - u^2 v^2 \sin 2 \alpha
    \end{pmatrix},
}
with 
\eq{
    w = v^2 - u^2.
}

\subsection{Combined elliptical and linear polarization fields}
\label{sec:app1_ellip_pi}

The Hamiltonian for two molecules under combined elliptical and linear polarization fields is
\eq{
    H = \mathcal{E} - 8 \sqrt{\frac{2}{15}} \pi^{3 / 2} \frac{d_0^2}{4 \pi \epsilon_0 r^3} \sum_{m = - 2}^{2} Y_{2, m}^*(\vu*{r}) \Sigma_{2, m},
}
where 
\eq{\mathcal{E} = 
    \begin{pmatrix}
        \delta + \Omega_{\text{eff}} & 0 & 0 & 0 & 0 & 0 & 0 \\
        0 & \frac{1}{2} (\delta + \Omega_{\text{eff}}) & 0 & 0 & 0 & 0 & 0 \\
        0 & 0 & \frac{1}{2} (\delta + \Omega_{\text{eff}}) & 0 & 0 & 0 & 0 \\
        0 & 0 & 0 & \delta & 0 & 0 & 0 \\
        0 & 0 & 0 & 0 & \frac{1}{2} (\delta - \Omega_{\text{eff}}) & 0 & 0 \\
        0 & 0 & 0 & 0 & 0 & \frac{1}{2} (\delta - \Omega_{\text{eff}}) & 0 \\
        0 & 0 & 0 & 0 & 0 & 0 & \delta - \Omega_{\text{eff}}
    \end{pmatrix},
}
\eq{\Sigma_{2, 0} = \frac{1}{4 \pi \sqrt{6}}
    \begin{pmatrix}
        -2 u^2 v^2  w_1 & 3 \sqrt{2} u v^2 p q & 0 & \sqrt{2} u v w w_1 & 3 \sqrt{2} u^2 v p q & 0 & 2 u^2 v^2 w_1 \\
        3 \sqrt{2} u v^2 p q & 2 v^2 w_2 & 0 & - 3 v w p q & 2 u v w_2 & 0 & - 3 \sqrt{2} u v^2 p q \\
        0 & 0 & -v^2 & 0 & 0 & -u v & 0 \\
        \sqrt{2} u v w w_1 & - 3 v w p q & 0 & - w^2 w_1 & - 3 u w p q & 0 & - \sqrt{2} u v w w_1 \\
        3 \sqrt{2} u^2 v p q & 2 u v w_2 & 0 & - 3 u w p q & 2 u^2 w_2 & 0 & - 3 \sqrt{2} u^2 v p q \\
        0 & 0 & -u v & 0 & 0 & -u^2 & 0 \\
        2 u^2 v^2 w_1 & - 3 \sqrt{2} u v^2 p q & 0 & - \sqrt{2} u v w w_1 & - 3 \sqrt{2} u^2 v p q & 0 & -2 u^2 v^2 w_1 
    \end{pmatrix},
}
{\footnotesize\eq{
    \Sigma_{2, 1} = \frac{1}{4 \pi \sqrt{2}}
    \begin{pmatrix}
        2 u^2 v^2 p q w_3 & \sqrt{2} u v^2 w_4 & - \sqrt{2} u v^2 q \cos\alpha & - \sqrt{2} u v w p q w_3 & \sqrt{2} u^2 v w_4 & - \sqrt{2} u^2 v q \cos\alpha & - 2 u^2 v^2 p q w_3  \\
        - \sqrt{2} u v^2 w_5 & - v^2 p q w_3 & - v^2 p \cos\alpha & v w w_5 & - u v p q w_3 & - u v p \cos\alpha & \sqrt{2} u v^2 w_5 \\
        - \sqrt{2} u v^2 q \sin\alpha & - v^2 p \sin\alpha & 0 & v w q \sin\alpha & - u v p \sin\alpha & 0 & \sqrt{2} u v^2 q \sin\alpha \\
        - \sqrt{2} u v w p q w_3 & - v w w_4 & v w q \cos\alpha & w^2 p q w_3 & - u w w_4 & u w q \cos\alpha  & \sqrt{2} u v w p q w_3 \\
        - \sqrt{2} u^2 v w_5 & - u v p q w_3 & - u v p \cos\alpha & u w w_5 & - u^2 p q w_3 & - u^2 p \cos\alpha & \sqrt{2} u^2 v w_5 \\
        - \sqrt{2} u^2 v q \sin\alpha & - u v p \sin\alpha & 0 & u w q \sin\alpha & - u^2 p \sin\alpha & 0 & \sqrt{2} u^2 v q \sin\alpha \\
        - 2 u^2 v^2 p q w_3 & - \sqrt{2} u v^2 w_4 & \sqrt{2} u v^2 q \cos\alpha & \sqrt{2} u v w p q w_3 & - \sqrt{2} u^2 v w_4 & \sqrt{2} u^2 v q \cos\alpha & 2 u^2v^2 p q w_3
    \end{pmatrix},
}}
{\footnotesize\eq{
    \Sigma_{2, 2} = - \frac{1}{4 \pi}
    \begin{pmatrix}
        -u^2 v^2 p^2 \sin 2 \alpha & \frac{\sqrt{2}}{2} u v^2 p q \sin 2 \alpha & - \sqrt{2} u v^2 p \cos^2 \alpha & \frac{\sqrt{2}}{2} u v w p^2 \sin 2 \alpha & \frac{\sqrt{2}}{2} u^2 v p q \sin 2 \alpha & -\sqrt{2} u^2 v p \cos^2 \alpha & u^2 v^2 p^2 \sin 2 \alpha \\
        \frac{\sqrt{2}}{2} u v^2 p q \sin 2 \alpha & - \frac{1}{2} v^2 q^2 \sin 2 \alpha & v^2 q \cos^2 \alpha & - \frac{1}{2} v w p q \sin 2 \alpha & - \frac{1}{2} u v q^2 \sin 2 \alpha & u v q \cos^2 \alpha & - \frac{\sqrt{2}}{2} u v^2 p q \sin 2 \alpha \\
        \sqrt{2} u v^2 p \sin^2 \alpha & - v^2 q \sin^2 \alpha & \frac{1}{2} v^2 \sin 2 \alpha & - v w p \sin^2 \alpha & - u v q \sin^2 \alpha & \frac{1}{2} u v \sin 2 \alpha & - \sqrt{2} u v^2 p \sin^2 \alpha \\
        \frac{\sqrt{2}}{2} u v w p^2 \sin 2 \alpha & - \frac{1}{2} v w p q \sin 2 \alpha & v w p \cos^2 \alpha & - \frac{1}{2} w^2 p^2 \sin 2 \alpha & - \frac{1}{2} u w p q \sin 2 \alpha & u w p \cos^2 \alpha & - \frac{\sqrt{2}}{2} u v w p^2 \sin 2 \alpha \\
        \frac{\sqrt{2}}{2} u^2 v p q \sin 2 \alpha & - \frac{1}{2} u v q^2 \sin 2 \alpha & u v q \cos^2 \alpha & - \frac{1}{2} u w p q \sin 2 \alpha & - \frac{1}{2} u^2 q^2 \sin 2 \alpha & u^2 q \cos^2 \alpha & - \frac{\sqrt{2}}{2} u^2 v p q \sin 2 \alpha \\
        \sqrt{2} u^2 v p \sin^2 \alpha & - u v q \sin^2 \alpha & \frac{1}{2} u v \sin 2 \alpha & - u w p \sin^2 \alpha & - u^2 q \sin^2 \alpha & \frac{1}{2} u^2 \sin 2 \alpha & - \sqrt{2} u^2 v p \sin^2 \alpha \\
        u^2 v^2 p^2 \sin 2 \alpha & - \frac{\sqrt{2}}{2} u v^2 p q \sin 2 \alpha & \sqrt{2} u v^2 p \cos^2 \alpha & - \frac{\sqrt{2}}{2} u v w p^2 \sin 2 \alpha & - \frac{\sqrt{2}}{2} u^2 v p q \sin 2 \alpha & \sqrt{2} u^2 v p \cos^2 \alpha & - u^2 v^2 p^2 \sin 2 \alpha
    \end{pmatrix},
}}
with 
\eq{
    w &= v^2 - u^2, w_1 = p^2 - 2 q^2, w_2 = p^2 - q^2 / 2, \\
    w_3 &= \cos\alpha - \sin\alpha, w_4 = p^2 \cos\alpha + q^2 \sin\alpha, w_5 = p^2 \sin\alpha + q^2 \cos\alpha.
}

The coefficients $c_1$ and $c_2$ in the main text are given as
\eq{
    c_1 &= \qty(1 + \mathcal{F}) \qty[\qty(1 - \mathcal{F}) \sin^2 \theta - \sqrt{2} s \mathcal{G} \sin 2 \theta + 2 s^2 \cos^2 \theta] \sin^2 \theta, \\
    c_2 &= \frac{1}{2} \qty[\qty(s^4 + 1) \qty(1 - \mathcal{F}) + 2 s^2 (\sin 2 \alpha - \cos 2 \phi)] \sin^2 2 \theta
    + 9 s^2 \cos^4 \theta + 3 \sqrt{2} s \qty(s^2 - 1) \mathcal{G} \cos^2 \theta \sin 2 \theta \\
    &+ s^2 \qty(1 - \mathcal{F} \sin^2 \theta)^2 - 6 s^2 \qty(1 - \mathcal{F} \sin^2 \theta) \cos^2 \theta
    + \sqrt{2} s \qty(1 - s^2) \mathcal{G} \qty(1 - \mathcal{F} \sin^2 \theta) \sin 2 \theta.
}

\subsection{Combined elliptical polarization and static fields}
\label{sec:app1_ellip_static}

The Hamiltonian for two molecules under combined elliptical polarization and static fields is 
\eq{
    H = \mathcal{E} - 8 \sqrt{\frac{2}{15}} \pi^{3 / 2} \frac{d_0^2}{4 \pi \epsilon_0 r^3} \sum_{m = - 2}^{2} Y_{2, m}^*(\vu*{r}) \Sigma_{2, m},
}
where 
\eq{\mathcal{E} = 
    \begin{pmatrix}
        \delta + \Omega_{\text{eff}} & 0 & 0 & 0 & 0 \\
        0 & \frac{1}{2} (\delta + \Omega_{\text{eff}}) & 0 & 0 & 0 \\
        0 & 0 & \delta & 0 & 0 \\
        0 & 0 & 0 & \frac{\delta - \Omega_{\text{eff}}}{2} & 0 \\
        0 & 0 & 0 & 0 & \delta - \Omega_{\text{eff}}
    \end{pmatrix},
}
{\small\eq{\Sigma_{2, 0} = \frac{1}{4 \pi \sqrt{6}} q^2
    \begin{pmatrix}
        2 v^2 (- u^2 + 4 v^2 p^2) & 0 & \sqrt{2} u v (- u^2 + v^2 (1 + 8 p^2)) & 0 & 2 u^2 v^2 (1 + 4 p^2) \\
        0 & - v^2 & 0 & - u v & 0 \\
        \sqrt{2} u v (- u^2 + v^2 (1 + 8 p^2)) & 0 & - (u^4 + v^4 - 2 u^2 v^2 (1 + 8 p^2)) & 0 & \sqrt{2} u v (- v^2 + u^2 (1 + 8 p^2)) \\
        0 & - u v & 0 & - u^2 & 0 \\
        2 u^2 v^2 (1 + 4 p^2) & 0 & \sqrt{2} u v (- v^2 + u^2 (1 + 8 p^2)) & 0 & 2 u^2 (- v^2 + 4 u^2 p^2)
    \end{pmatrix},
}}
\eq{
    \Sigma_{2, 2} = - \frac{1}{4 \pi} q^2
    \begin{pmatrix}
        u^2 v^2 \sin 2 \alpha & \sqrt{2} u v^2 \cos^2 \alpha & - \frac{\sqrt{2}}{2} u v w \sin 2 \alpha & \sqrt{2} u^2 v \cos^2 \alpha & - u^2 v^2 \sin 2 \alpha \\
        - \sqrt{2} u v^2 \sin^2 \alpha & - \frac{1}{2} v^2 \sin 2 \alpha & v w \sin^2 \alpha & - \frac{1}{2} u v \sin 2 \alpha & \sqrt{2} u v^2 \sin^2 \alpha \\
        - \frac{\sqrt{2}}{2} u v w \sin 2 \alpha & - v w \cos^2 \alpha & \frac{1}{2} w^2 \sin 2 \alpha & - u w \cos^2 \alpha & \frac{\sqrt{2}}{2} u v w \sin 2 \alpha \\
        - \sqrt{2} u^2 v \sin^2 \alpha & - \frac{1}{2} u v \sin 2 \alpha & u w \sin^2 \alpha & - \frac{1}{2} u^2 q^2 \sin 2 \alpha & \sqrt{2} u^2 v \sin^2 \alpha \\
        - u^2 v^2 \sin 2 \alpha & - \sqrt{2} u v^2 \cos^2 \alpha & \frac{\sqrt{2}}{2} u v w \sin 2 \alpha & - \sqrt{2} u^2 v \cos^2 \alpha & u^2 v^2 \sin 2 \alpha
    \end{pmatrix},
}

The coefficients $c_1$ and $c_2$ in the main text are given as
\eq{
    c_1 &= \qty[\cos^4 \alpha + \sin^4 \alpha - \frac{1}{2} \cos 4 \alpha \sin^2 2 \alpha] \sin^4 \theta, \\
    c_2 &= \qty[(1 - s^2 (1 + 8 p^2)) (3 \cos^2 \theta - 1) + 3 (1 - s^2) \mathcal{F} \sin^2 \theta]^2.
}

\section{Scattering length for the multi-channel scattering process}
\label{sec:app2}

The Schrödinger equation for the multi-channel scattering process is 
\eq{
    \sum_{\nu'} \qty( - \frac{\laplacian}{2 M} \delta_{\nu \nu'} + V_{\nu \nu'} ) \psi_{\nu'} = \frac{k_\nu^2}{2 M} \psi_\nu,
}
where $M$ is the reduced mass of two molecules, $k_\nu^2 / (2 M)$ is the incident energy of the $\nu$-th channel, and $k_\nu = \sqrt{k_1^2 - 2 M \mathcal{E}_\nu}$. Hereafter, we set $\hbar = 1$. Furthermore, the wavefunction $\psi_\nu$ can be expanded as $\psi_\nu = \sum_{l m} r^{-1} \phi_{\nu l m} Y_{l m}$. Then the Schrödinger equation becomes
\eq{
    \sum_{\nu' l' m'} \qty[ - \frac{1}{2 M} \qty( \frac{\partial^2}{\partial r^2} - \frac{l (l + 1)}{r^2} ) \delta_{\nu \nu'} \delta_{l l'} \delta_{m m'} - \frac{d_0^2}{4 \pi \epsilon_0 r^3} \sum_{s} (\varLambda_{s})_{\nu l m, \nu' l' m'} ] \phi_{\nu' l' m'} = \frac{k_\nu^2}{2 M} \phi_{\nu l m},
}
where the potential matrix element is 
\eq{
    (\varLambda_{s})_{\nu l m, \nu' l' m'} = 4 \pi \sqrt{\frac{2 (2 l' + 1)}{3 (2 l + 1)}} C^{l 0}_{l' 0 2 0} C^{l m}_{l' m' 2 s} \Sigma_{2 s}^\dagger,
}
with the Clebsch-Gordan coefficient $C_{l' m' l_1 m_1}^{l m}$. For the numerical calculation, we write the dimensionless Schrödinger equation as
\eq{
    \sum_{l'} \qty[- \qty( \frac{\partial^2}{\partial x^2} - \frac{ l (l + 1)}{x^2} ) \delta_{\nu \nu'} \delta_{l l'} \delta_{m m'} - \frac{2 M a_B^2}{\hbar^2} \frac{d^2}{4 \pi \epsilon_0 r^3} \sum_{s} (\varLambda_{s})_{\nu l m, \nu' l' m'} ] \phi_{\nu' l' m'} = a_B^2 k_\nu^2 \phi_{\nu l m}.
}
Finally, by defining 
\eq{
    \mathcal{V} = \qty[a_B^2 k_\nu^2 - \frac{l (l + 1)}{x^2}] \delta_{\nu \nu'} \delta_{ll'} \delta_{mm'} + \frac{2 M a_B^2}{\hbar^2} \frac{d^2}{4 \pi \epsilon_0 r^3} \sum_{s} (\varLambda_{s})_{\nu l m, \nu' l' m'},
}
the dimensionless Schrödinger equation becomes
\eq{
    \qty[\pdv[2]{x} + \mathcal{V}] \phi = 0.
}

To numerically solve the above equation, we set the boundary condition as $\phi(0) = 0$, and $\phi'(0)$ is arbitrary as long as it leads to linearly independent solutions. A convenient choice is $\phi'(0) = 1$. On the other hand, for the incident particles, the asymptotic wavefunction at $r \rightarrow \infty$ is 
\eq{
    \phi = \vb{J} + \vb{N} \vb{K},
}
where $\vb{J}$ and $\vb{N}$ are diagonal matrices, whose elements are made up of Riccati-Bessel functions:
\eq{
    \vb{J}_l &= k_\nu^{-1/2} \hat{j}_{l}(k_\nu x), \\
    \vb{N}_l &= k_\nu^{-1/2} \hat{n}_{l}(k_\nu x).
}
Here, $\hat{j}_{l}(z) = z j_l(z)$ and $\hat{n}_l(z) = - z n_l(z)$, where $j_l(z)$ and $n_l(z)$ are the spherical Bessel functions. $k_\nu$ is the channel wave number.

For convenience, we define the matrix $\mathcal{Y} = (\partial_x \phi) \phi^{-1}$, combining the boundary conditions to solve the dimensionless Schrödinger equation using the log-derivative method, which satisfies
\eq{
    \partial_x \mathcal{Y} = - \mathcal{V} - \mathcal{Y}^2.
}
We then obtain the reaction matrix $\vb{K}$ and the scattering matrix $\vb{S}$ as
\eq{
    \vb{K} &= - \qty[\mathcal{Y} \vb{N} - \vb{N}']^{-1} \qty[\mathcal{Y} \vb{J} - \vb{J}'], \\
    \vb{S} &= \qty(I - i \vb{K})^{-1} \qty(I + i \vb{K}).
}
Finally, the $t$ matrix and the scattering length can be calculated as 
\eq{
    t &= \frac{1}{k_1} \frac{\vb{K}}{1 - i \vb{K}}, \\
    a_{\nu l m, \nu' l' m'} &= \lim_{k_1 \rightarrow 0} - \frac{\vb{K}_{\nu l m, \nu' l' m'}}{k_1} = - \lim_{k_1 \rightarrow 0} t_{1 0 0, 1 0 0}.
}

\section{Scattering length for the single-channel scattering process}
\label{sec:app3}

To validate the effective potential, we compare the scattering length obtained from the effective potential with that from the full Hamiltonian. The single-channel scattering process is described by the Schrödinger equation:
\eq{
    \qty[- \frac{\laplacian}{2 M} + V_{\text{eff}}] \psi_{++} = \frac{k_1^2}{2 M} \psi_{++}.
}
Similarly, in the partial wave expansion, the Schrödinger equation becomes
\eq{
    \sum_{l' m'} \qty[ - \frac{1}{2 M} \qty(\pdv[2]{r} - \frac{l (l + 1)}{r^2}) \delta_{l l'} \delta_{m m'} + \qty(V_{\text{eff}})_{l m, l' m'} ] \phi_{l' m'} = \frac{k_1^2}{2 M} \phi_{l m}.
}

\subsection{Elliptical polarization field}
\label{sec:app3_ellip}

According to Eq.~\eqref{eq:eff_pot_ellip}, the effective potential matrix elements are
\eq{
    \qty(V_{\text{eff}})_{l m, l' m'} &= \frac{C_3}{r^3} \sqrt{\frac{2 l' + 1}{2 l + 1}} C^{l 0}_{l' 0 2 0} \qty[2 C^{l m}_{l' m' 2 0} + \sqrt{6} \sin 2 \alpha \qty(C^{l m}_{l' m' 2 2} + C^{l m}_{l' m' 2 -2})] \\
    &+ \frac{C_6}{r^6} \sqrt{\frac{2 l' + 1}{2 l + 1}} \qty[\frac{4}{5} \qty(1 - \frac{1}{3} \sin^2 2 \alpha)] \delta_{l l'} \delta_{m m'} \\
    &- \frac{C_6}{r^6} \sqrt{\frac{2 l' + 1}{2 l + 1}} C^{l 0}_{l' 0 2 0} \qty[\frac{4}{7} \qty(1 - \frac{2}{3} \sin^2 2 \alpha) C^{l m}_{l' m' 2 0} + \frac{2}{7} \sqrt{\frac{2}{3}} \sin 2 \alpha \qty(C^{l m}_{l' m' 2 2} + C^{l m}_{l' m' 2 -2})] \\
    &- \frac{C_6}{r^6} \sqrt{\frac{2 l' + 1}{2 l + 1}} C^{l 0}_{l' 0 4 0} \qty[\frac{4}{35} \qty(2 + \sin^2 2 \alpha) C^{l m}_{l' m' 4 0} + \frac{4}{7} \sqrt{\frac{2}{5}} \sin 2 \alpha \qty(C^{l m}_{l' m' 4 2} + C^{l m}_{l' m' 4 -2})] \\
    &- \frac{C_6}{r^6} \sqrt{\frac{2 l' + 1}{2 l + 1}} C^{l 0}_{l' 0 4 0} \qty[2 \sqrt{\frac{2}{35}} \qty(C^{l m}_{l' m' 4 4} + C^{l m}_{l' m' 4 -4})].
}

\subsection{Combined elliptical and linear polarization fields}
\label{sec:app3_ellip_pi}

According to the equation in Eq.~\eqref{eq:eff_pot_ellip_pi}, the effective potential matrix element is
{\footnotesize\be    
&\qty(V_{\text{eff}})_{lm, l'm'} \\
&= \frac{C_3}{r^3} \sqrt{\frac{2 l' + 1}{2 l + 1}} C^{l 0}_{l' 0 2 0} \qty[2 \qty(1 - 2 s^2) C^{l m}_{l' m' 2 0} + 2 \sqrt{3} s (- \cos \alpha + \sin \alpha) \qty(C^{l m}_{l' m' 2 1} - C^{l m}_{l' m' 2 - 1}) + \sqrt{6} \sin 2 \alpha \qty(C^{l m}_{l' m' 2 2} + C^{l m}_{l' m' 2 - 2})] \\
&+ \frac{C_6}{r^6} \sqrt{\frac{2 l' + 1}{2 l + 1}} \qty[\frac{2}{15} \qty(3 + 2 s^2 + \cos 4 \alpha ) + p^2 \frac{4}{15} \qty(1 + s^4 + 2 s^2 \sin 2 \alpha ) + p^2 \frac{9}{5} s^2 - p^2 \frac{1}{15} s^2 (13 + 2 \cos 4 \alpha)] \delta_{ll'} \delta_{mm'} \\
&+ \frac{C_6}{r^6} \sqrt{\frac{2 l' + 1}{2 l + 1}} C^{l 0}_{l' 0 2 0} \qty[\frac{4}{21} \qty(- 3 + s^2 - \cos 4 \alpha ) + p^2 \frac{4}{21} \qty(1 + s^4 + 2 s^2 \sin 2 \alpha) + p^2 \frac{36}{7} s^2 + p^2 \frac{4}{21} s^2 (- 22 + \cos 4 \alpha)] C^{l m}_{l' m' 2 0} \\
&+ \frac{C_6}{r^6} \sqrt{\frac{2 l' + 1}{2 l + 1}} C^{l 0}_{l' 0 2 0} \qty[\frac{2}{7 \sqrt{3}} s (3 \cos \alpha + \cos 3 \alpha - 3 \sin \alpha + \sin 3 \alpha) + p^2 \frac{6 \sqrt{3}}{7} s \qty(\cos \alpha - s^2 \cos \alpha - \sin \alpha + s^2 \sin \alpha)] \qty(C^{l m}_{l' m' 2 1} - C^{l m}_{l' m' 2 - 1}) \\
&+ \frac{C_6}{r^6} \sqrt{\frac{2 l' + 1}{2 l + 1}} C^{l 0}_{l' 0 2 0} \qty[p^2 \frac{2}{7 \sqrt{3}} s \qty(- 8 \cos \alpha + 8 s^2 \cos \alpha + \cos 3 \alpha - s^2 \cos 3 \alpha + 8 \sin \alpha - 8 s^2 \sin \alpha + \sin 3 \alpha - s^2 \sin 3 \alpha)] \qty(C^{l m}_{l' m' 2 1} - C^{l m}_{l' m' 2 - 1}) \\
&+ \frac{C_6}{r^6} \sqrt{\frac{2 l' + 1}{2 l + 1}} C^{l 0}_{l' 0 2 0} \qty[\frac{2}{7} \sqrt{\frac{2}{3}} s^2 \sin 2 \alpha - p^2 \frac{2}{7} \sqrt{\frac{2}{3}} \qty(2 s^2 + \sin 2 \alpha + s^4 \sin 2 \alpha) - p^2 \frac{8}{7} \sqrt{\frac{2}{3}} s^2 \sin 2 \alpha] \qty(C^{l m}_{l' m' 2 2} + C^{l m}_{l' m' 2 - 2}) \\
&+ \frac{C_6}{r^6} \sqrt{\frac{2 l' + 1}{2 l + 1}} C^{l 0}_{l' 0 4 0} \qty[\frac{2}{35} \qty(3 - 8 s^2 + \cos 4 \alpha) - p^2 \frac{16}{35} \qty(1 + s^4 + 2 s^2 \sin 2 \alpha) + p^2 \frac{72}{35} s^2 + p^2 \frac{2}{35} s^2 \qty(1 - \cos 4 \alpha)] C^{l m}_{l' m' 4 0} \\
&+ \frac{C_6}{r^6} \sqrt{\frac{2 l' + 1}{2 l + 1}} C^{l 0}_{l' 0 4 0} \qty[\frac{\sqrt{2}}{7 \sqrt{5}} s (- 3 \cos \alpha - \cos 3 \alpha + 3 \sin \alpha - \sin 3 \alpha) + p^2 \frac{12}{7} \sqrt{\frac{2}{5}} s (\cos \alpha - s^2 \cos \alpha - \sin \alpha + s^2 \sin \alpha)] \qty(C^{l m}_{l' m' 4 1} - C^{l m}_{l' m' 4 - 1}) \\
&+ \frac{C_6}{r^6} \sqrt{\frac{2 l' + 1}{2 l + 1}} C^{l 0}_{l' 0 4 0} \qty[p^2 \frac{1}{7} \sqrt{\frac{2}{5}} s \qty(\cos \alpha - s^2 \cos \alpha - \cos 3 \alpha + s^2 \cos 3 \alpha - \sin \alpha + s^2 \sin \alpha - \sin 3 \alpha + s^2 \sin 3 \alpha)] \qty(C^{l m}_{l' m' 4 1} - C^{l m}_{l' m' 4 - 1}) \\
&+ \frac{C_6}{r^6} \sqrt{\frac{2 l' + 1}{2 l + 1}} C^{l 0}_{l' 0 4 0} \qty[\frac{4}{7} \sqrt{\frac{2}{5}} s^2 \sin 2 \alpha - p^2 \frac{4}{7} \sqrt{\frac{2}{5}} \qty(2 s^2 + \sin 2 \alpha + s^4 \sin 2 \alpha) + p^2 \frac{12}{7} \sqrt{\frac{2}{5}} s^2 \sin 2 \alpha] \qty(C^{l m}_{l' m' 4 2} + C^{l m}_{l' m' 4 - 2}) \\
&+ \frac{C_6}{r^6} \sqrt{\frac{2 l' + 1}{2 l + 1}} C^{l 0}_{l' 0 4 0} \qty[\sqrt{\frac{2}{35}} s (- \cos \alpha + \cos 3 \alpha + \sin \alpha + \sin 3 \alpha)] \qty(C^{l m}_{l' m' 4 3} - C^{l m}_{l' m' 4 - 3}) \\
&+ \frac{C_6}{r^6} \sqrt{\frac{2 l' + 1}{2 l + 1}} C^{l 0}_{l' 0 4 0} \qty[p^2 \sqrt{\frac{2}{35}} s \qty(- \cos \alpha + s^2 \cos \alpha + \cos 3 \alpha - s^2 \cos 3 \alpha + \sin \alpha - s^2 \sin \alpha + \sin 3 \alpha - s^2 \sin 3 \alpha)] \qty(C^{l m}_{l' m' 4 3} - C^{l m}_{l' m' 4 - 3}) \\
&+ \frac{C_6}{r^6} \sqrt{\frac{2 l' + 1}{2 l + 1}} C^{l 0}_{l' 0 4 0} \qty[\sqrt{\frac{2}{35}} (- 1 + \cos 4 \alpha) + p^2 \sqrt{\frac{2}{35}} s^2 (1 - \cos 4 \alpha)] \qty(C^{l m}_{l' m' 4 4} + C^{l m}_{l' m' 4 - 4}).
\ee}

\subsection{Combined elliptical polarization and static fields}
\label{sec:app3_ellip_static}

According to the equation in Eq.~\eqref{eq:eff_pot_ellip_static}, the effective potential matrix element is
{\footnotesize\eq{
    &\qty(V_\text{eff})_{ll'mm'} \\
    &= \frac{C_3}{r^3} \sqrt{\frac{2 l' + 1}{2 l + 1}} C^{l 0}_{l' 0 2 0} \qty[2 (1 - 4 s^2 p^2) C^{l m}_{l' m' 2 0} + \sqrt{6} \sin(2 \alpha) \qty(C^{l m}_{l' m' 2 2} + C^{l m}_{l' m' 2 - 2})] \\
    &+ \frac{C_6}{r^6} \sqrt{\frac{2 l' + 1}{2 l + 1}} \qty[\frac{2}{15} \qty(4 - \cos 4 \alpha + \cos 8 \alpha) + \frac{u^4}{9} \frac{2}{5} \qty(5 - 10 s^2 - 32 s^2 p^2 + 5 s^4 + 32 s^4 p^2 + 128 s^4 p^4 - 3 \cos 4 \alpha + 6 s^2 \cos 4 \alpha - 3 s^4 \cos 4 \alpha)] \delta_{ll'} \delta_{mm'} \\
    &+ \frac{C_6}{r^6} \sqrt{\frac{2 l' + 1}{2 l + 1}} C^{l 0}_{l' 0 2 0} \qty[\frac{4}{21} \qty(- 4 + \cos 4 \alpha - \cos 8 \alpha) + \frac{u^4}{9} \frac{4}{7} \qty(- 1 + 2 s^2 - 32 s^2 p^2 - s^4 + 32 s^4 p^2 + 128 s^4 p^4 + 3 \cos 4 \alpha - 6 s^2 \cos 4 \alpha + 3 s^4 \cos 4 \alpha)] C^{l m}_{l' m' 2 0} \\
    &+ \frac{C_6}{r^6} \sqrt{\frac{2 l' + 1}{2 l + 1}} C^{l 0}_{l' 0 2 0} \qty[\frac{u^4}{9} \frac{8 \sqrt{6}}{7} \qty(- 1 + 2 s^2 + 8 s^2 p^2 - s^4 - 8 s^4 p^2) \sin 2 \alpha] \qty(C^{l m}_{l' m' 2 2} + C^{l m}_{l' m' 2 - 2}) \\
    &+ \frac{C_6}{r^6} \sqrt{\frac{2 l' + 1}{2 l + 1}} C^{l 0}_{l' 0 4 0} \qty[\frac{2}{35} \qty(4 - \cos 4 \alpha + \cos 8 \alpha) + \frac{u^4}{9} \frac{18}{35} \qty(5 - 10 s^2 - 64 s^2 p^2 + 5 s^4 + 64s^4 p^2 + 256 s^4 p^4 - \cos 4 \alpha + 2 s^2 \cos 4 \alpha - s^4 \cos 4 \alpha)] C^{l m}_{l' m' 4 0} \\
    &+ \frac{C_6}{r^6} \sqrt{\frac{2 l' + 1}{2 l + 1}} C^{l 0}_{l' 0 4 0} \qty[\frac{u^4}{9} \frac{36}{7} \sqrt{\frac{2}{5}} \qty(1 - 2 s^2 - 8 s^2 p^2 + s^4 + 8 s^4 p^2) \sin 2 \alpha] \qty(C^{l m}_{l' m' 4 2} + C^{l m}_{l' m' 4 - 2}) \\
    &+ \frac{C_6}{r^6} \sqrt{\frac{2 l' + 1}{2 l + 1}} C^{l 0}_{l' 0 4 0} \qty[\frac{u^4}{9} 9 \sqrt{\frac{2}{35}} \qty(1 - 2 s^2 + s^4 - \cos 4 \alpha + 2 s^2 \cos 4 \alpha - s^4 \cos 4 \alpha)] \qty(C^{l m}_{l' m' 4 4} + C^{l m}_{l' m' 4 - 4}).
}}

\section{Scattering length for the effective potential under the Born approximation}
\label{sec:app4}

Because the shielding core decays rapidly, the effective potential can be approximated as a zero-range contact interaction supplemented by a long-range dipole-dipole interaction. Thus, we can write the effective pseudopotential as
\eq{
    V_{\text{eff}} = u_0 \delta(\vb*{r}) + \sum_{l_1, m_1} \gamma_{l_1, m_1} \frac{Y_{l_1, m_1}(\vu*{r})}{r^3},
}
whose first Born amplitude is
\eq{
    f(\vb*{k}', \vb*{k}) = - \frac{M}{2 \pi \hbar^2} \int \dd[3]{\vb*{r}}\, e^{i (\vb*{k} - \vb*{k}') \cdot \vb*{r}} V_{\text{eff}}(\vb*{r}).
}
Expanding each plane wave in spherical harmonics, we have
\eq{
    f(\vb*{k}', \vb*{k}) = - \frac{M}{2 \pi \hbar^2} \int \dd[3]{\vb*{r}}\, V_{\text{eff}}(\vb*{r}) \sum_{l', m'} 4 \pi (-i)^{l'} j_{l'}(k' r) Y_{l', m'}^*(\vu*{r}) Y_{l', m'}(\vu*{k}') \sum_{l, m} 4 \pi i^l j_l(k r) Y_{l, m}(\vu*{r}) Y_{l, m}^*(\vu*{k}).
}
Separating the radial and angular parts, the above equation becomes
\eq{
    f(\vb*{k}', \vb*{k}) &= - \frac{M}{2 \pi \hbar^2} u_0 (4 \pi)^2 \left(\frac{1}{\sqrt{4 \pi}}\right)^2 Y_{0, 0}(\vu*{k}') Y_{0, 0}^*(\vu*{k}) \\
    &\quad - \frac{M}{2 \pi \hbar^2} \sum_{l_1, m_1} \gamma_{l_1, m_1} (4 \pi)^2 \sum_{l, m} \sum_{l', m'} \mathcal{T}_{l, m}^{l', m'}(l_1, m_1) Y_{l, m}^*(\vu*{k}) Y_{l', m'}(\vu*{k}'),
}
where
\eq{
    \mathcal{T}_{l, m}^{l', m'}(l_1, m_1) &= i^{l + l'} \mathcal{R}_l^{l'} \mathcal{I}_{l m}^{l' m'}(l_1, m_1), \\
    \mathcal{I}_{l m}^{l' m'}(l_1, m_1) &= \int \dd\Omega\, Y_{l', m'}^*(\vu*{r}) Y_{l_1, m_1}(\vu*{r}) Y_{l, m}(\vu*{r}) = \sqrt{\frac{(2l+1)(2l_1+1)}{4\pi(2l'+1)}} C^{l' 0}_{l 0\, l_1 0} C^{l' m'}_{l m\, l_1 m_1}, \\
    \mathcal{R}_l^{l'} &= \int_0^\infty \dd{r}\, r^2 \frac{j_{l'}(k' r) j_l(k r)}{r^3} = \frac{\pi}{8} \frac{\Gamma\left(\frac{l + l'}{2}\right)}{\Gamma\left(\frac{l - l' + 3}{2}\right) \Gamma\left(\frac{l' - l + 3}{2}\right) \Gamma\left(\frac{l + l' + 4}{2}\right)}, \\
    u_0 &= \frac{4 \pi \hbar^2 a_s}{2 M}.
}
Finally, the scattering amplitude can be written as
\eq{
    f(\vb*{k}', \vb*{k}) = - 4 \pi a_s Y_{0, 0}(\vu*{k}') Y_{0, 0}^*(\vu*{k}) + 4 \pi \sum_{l, m, l', m'} t_{l m}^{l' m'} Y_{l, m}^*(\vu*{k}) Y_{l', m'}(\vu*{k}'),
}
where
\eq{
    t_{l m}^{l' m'} = - \frac{M}{2 \pi \hbar^2} \sum_{l_1, m_1} \gamma_{l_1, m_1} (4 \pi) \mathcal{T}_{l, m}^{l', m'}(l_1, m_1).
}
Obviously, the s-wave scattering length is $a_s = - t_{0 0}^{0 0}$.

\subsection{Elliptical polarization field}
\label{sec:app4_ellip}

The ratio $t_{l m}^{l' m'} / t_{0 0}^{2 0}$ for different $l, m, l', m'$ is shown in Tables~\ref{tab:scat_len_ellip_born} and~\ref{tab:scat_len_ellip_born_1}.

\begin{table}[h]
\begin{tabular}{ccccc}
    \hline\hline
    $(l \; m), (l' m')$ & $(0 \; 0)$ & $(2 \; 0)$ & $(4 \; 0)$ & $(6 \; 0)$ \\
    \hline
    $(0 \; 0)$ & $*$ & $1$ & $0$ & $0$ \\
    $(2 \; 0)$ & $1$ & $-0.63889$ & $0.14286$ & $0$ \\
    $(4 \; 0)$ & $0$ & $0.14286$ & $-0.17424$ & $0.05638$ \\
    $(6 \; 0)$ & $0$ & $0$ & $0.05638$ & $-0.08131$ \\
    \hline\hline
\end{tabular} 
\caption{The ratio $t_{l m}^{l' m'} / t_{0 0}^{2 0}$ for different $l, m, l', m'$. $\alpha = 0$.}
\label{tab:scat_len_ellip_born}  
\end{table}

\begin{table}[h]
\begin{tabular}{ccccc}
    \hline\hline
    $(l \; m), (l' m')$ & $(0 \quad \; \; 0)$ & $(2 \quad \; \; 0)$ & $(2 \; -2)$ & $(2 \quad \; \; 2)$ \\
    \hline
    $(0 \quad \; \; 0)$ & $*$ & $1$ & $-0.21268$ & $-0.21268$ \\
    $(2 \quad \; \; 0)$ & $1$ & $-0.63888$ & $-0.13587$ & $-0.13587$ \\
    $(2 \; -2)$ & $-0.21268$ & $-0.13587$ & $0.63888$ & $0$ \\
    $(2 \quad \; \; 2)$ & $-0.21268$ & $-0.13587$ & $0$ & $0.63888$ \\
    \hline\hline
\end{tabular}
\caption{The ratio $t_{l m}^{l' m'} / t_{0 0}^{2 0}$ for different $l, m, l', m'$. $\alpha = 5^\circ$.}
\label{tab:scat_len_ellip_born_1}
\end{table}

\subsection{Combined elliptical and linear polarization fields}
\label{sec:app4_ellip_pi}
The ratio $t_{l m}^{l' m'} / t_{0 0}^{2 0}$ for different $l, m, l', m'$ is shown in Tables~\ref{tab:scat_len_ellip_pi_born} and~\ref{tab:scat_len_ellip_pi_born_1}.

\begin{table}[h]
\begin{tabular}{ccccc}
    \hline\hline
    $(l \; m), (l' m')$ & $(0 \; 0)$ & $(2 \; 0)$ & $(2 \; 1)$ & $(2 \; 2)$ \\
    \hline
    $(0 \; 0)$ & $*$ & $1$ & $-0.37653$ & $0$ \\
    $(2 \; 0)$ & $1$ & $-0.63888$ & $0.12028$ & $0$ \\
    $(2 \; 1)$ & $-0.37653$ & $0.12028$ & $-0.31944$ & $0.29462$ \\
    $(2 \; 2)$ & $0$ & $0$ & $0.29462$ & $0.63888$ \\
    \hline\hline
\end{tabular} 
\caption{The ratio $t_{l m}^{l' m'} / t_{0 0}^{2 0}$ for different $l, m, l', m'$. $\alpha = 0$, $\Omega_\sigma = 15$ MHz, $\Omega_\pi = 3$ MHz.}
\label{tab:scat_len_ellip_pi_born}  
\end{table}

\begin{table}[h]
\begin{tabular}{ccccc}
    \hline\hline
    $(l \; m), (l' m')$ & $(0 \; 0)$ & $(2 \; 0)$ & $(2 \; 1)$ & $(2 \; 2)$ \\
    \hline
    $(0 \; 0)$ & $*$ & $1$ & $-2.14274$ & $0$ \\
    $(2 \; 0)$ & $1$ & $-0.63889$ & $0.68447$ & $0$ \\
    $(2 \; 1)$ & $-2.14274$ & $0.68447$ & $-0.31944$ & $1.67661$ \\
    $(2 \; 2)$ & $0$ & $0$ & $1.67661$ & $0.63888$ \\
    \hline\hline
\end{tabular}
\caption{The ratio $t_{l m}^{l' m'} / t_{0 0}^{2 0}$ for different $l, m, l', m'$. $\alpha = 0$, $\Omega_\sigma = 15$ MHz, $\Omega_\pi = 8$ MHz.}
\label{tab:scat_len_ellip_pi_born_1}
\end{table}

\subsection{Combined elliptical polarization and static fields}
\label{sec:app4_ellip_static}

The ratios $t_{l m}^{l' m'} / t_{0 0}^{2 0}$ for different $l, m, l', m'$ is shown in Table~\ref{tab:scat_len_ellip_static_born}.

\begin{table}[h]
    \begin{tabular}{ccccc}
        \hline\hline
        $(l \; m), (l' m')$ & $(0 \; 0)$ & $(2 \; 0)$ & $(4 \; 0)$ & $(6 \; 0)$ \\
        \hline
        $(0 \; 0)$ & $*$ & $1$ & $0$ & $0$ \\
        $(2 \; 0)$ & $1$ & $-0.63888$ & $0.14286$ & $0$ \\
        $(4 \; 0)$ & $0$ & $0.14286$ & $-0.17424$ & $0.05638$ \\
        $(6 \; 0)$ & $0$ & $0$ & $0.05638$ & $-0.08131$ \\
        \hline\hline
    \end{tabular}
    \caption{The ratio $t_{l m}^{l' m'} / t_{0 0}^{2 0}$ for different $l, m, l', m'$. $\alpha = 0$, $\delta = 10$~MHz, $\Omega_\sigma = 12$~MHz.}
    \label{tab:scat_len_ellip_static_born}
\end{table}

\end{widetext}

\normalem
% \bibliography{MSP.bib}

\end{document}